\documentclass[twocolumn,showpacs,superscriptaddress]{revtex4-1}

\usepackage{mathrsfs}
\usepackage{amsmath}
\usepackage{amssymb}
\usepackage{graphicx}
\usepackage{float}
\restylefloat{table}

\begin{document}

\title{Effect of a laser field in the confinement potential of two electrons in a double quantum dot}
\author{A. M. Maniero}
\email{angelo.maniero@ufob.edu.br}
\affiliation{Universidade Federal do Oeste da Bahia, 47808-021, Barreiras, BA, Brazil}

\author{C. R. de Carvalho}
\email{crenato@if.ufrj.br}
\affiliation{Instituto de F\'{\i}sica, Universidade Federal do Rio de Janeiro, Rio de Janeiro, 21941-972, RJ, Brazil}

\author{F. V. Prudente}
\email{prudente@ufba.br}
\affiliation{Universidade Federal do Oeste da Bahia, 47808-021, Barreiras, BA, Brazil}

\author{Ginette Jalbert}
\email{ginette@if.ufrj.br}
\affiliation{Instituto de F\'{\i}sica, Universidade Federal do Rio de Janeiro, Rio de Janeiro, 21941-972, RJ, Brazil}

\date{\today }
\begin{abstract}
 We have studied a system consisted of two coupled quantum dots containing two electrons subjected by a laser field. The effect of the laser is described by the dressed-band approach involving the concept of the conduction/valence effective mass, valid far from resonance. The interaction between the electrons and the quantum dots is described by a phenomenological tridimensional potential, which simulates  quantum dots in GaAs heterostructure. In this study we have employed the approach already presented in a previous work [Olavo {\em et al.}, J. Phys. B: At. Mol. Opt.
Phys. {\bf 49}, 145004 (2016)]. We have used a code based on the full interaction configuration method. We have employed as basis set the {\it Cartesian anisotropic Gaussian-type} orbitals which allows one to explore the confining characteristics of a potential due to their flexibility of using different exponents for each direction space. We present an analysis based on the energy levels of the singlet and triplet as function of the confinement parameters.
\end{abstract}


\maketitle

 \section{INTRODUCTION}
 
The advances of the experimental techniques used in semiconductor structures of nanoscopic scale \cite{heinzel-2010}  has increased the interest in the study of the physical properties of confined quantum systems. A consequence of this improvement on the manufacturing of semiconductor quantum dots (QDs) is the increase in the control of their size; this has attracted a great interest in the area of optoelectronics~\cite{H.Liu15}  and optical communications~\cite{Schmeckebier17}.
As long as the QD dimensions become of the order of nanometer, it has been noticed that the its physical properties are greatly affected by changes in its size \cite{Alivisatos96, ScienceSpecialIssue96}. Frequently the size and geometric form of the quantum dot has been treated in terms of confinement profile and strength \cite{Diercksen-JPB34-1987-01, Diercksen-JPB36-1681-03}. The influence of external fields on QDs has also attracted attention, in particular on double quantum dots (DQDs) aiming at quantum computation and general process in nanotechnology \cite{DiVincenzo-PRB59-2070-99, CRC-GJ-JAP94-2579-03, Szafran-PRB70-205318-04, Dybalski-PRB72-205432-05, Leburton-COSSMS10-114-06, Yamamoto-RPP76-092501-13,Prati-JPA48-065304-15}. The behavior of the exchange coupling ($J$), or exchange energy, has been one of the main subjects in the study of the properties of few-electron DQDs. In this context, different profiles of confining potential has been tried out such as  quartic \cite{DiVincenzo-PRB59-2070-99, CRC-GJ-JAP94-2579-03}, gaussian \cite{Leburton-COSSMS10-114-06, Leburton-JPCM21-095502-09} and few others \cite{Kwasniowski-JPCM20-215208-08, Pedersen-PRB81-193406-10}. In all these cases a two-dimension geometry has always been considered. 

In a previous work \cite{CRC-GJ-JAP94-2579-03} we have analyzed the exchange coupling ($J$) in the effective Heisenberg model within the Heitler-London approximation, so that it can be analytically calculated. We have discussed it as a function of the laser field and its detuning, as well as of the magnetic field. We have found that, due to the electronic confinement, the laser may play a role similar to the external magnetic field in the qualitative behavior of the exchange parameter ($J$). On the other hand, it has also been reported analytic expressions for the exchange coupling in 2D coupled quantum dots computed within the Heitler-London and the Hund-Mulliken approximations using different confining potentials under different regimes of magnetic field intensity \cite{Pedersen-PRB81-193406-10}.

Aiming more precise results, one finds a variety of numerical methods or techniques employed for calculations of the electronic structure of quantum systems such as atoms, ions and molecules confined by an external potential \cite{Diercksen-JPB36-1681-03, Diercksen-CPL349-215-01, Klobukowski-MP103-2599-05, Fred-JCP123-224701-05,LeSech-JPB45-205101-12, Bartkowiak-CP428-19-14, Sen14, JPB48-055002-15, Cruz-JPB50-135002-17}. The interest on this type of problem arose from the wide range of issues found in many branches of chemistry and physics \cite{Sabin-AQC57-58-09}. Naturally these methods are also suitable for the study of QDs since they can be seen as artificial atoms or molecules \cite{Sen14}.

In view of all these issues, we have developed a code \cite{olavo2016} which it allows to study arbitrary systems submitted to different confining potentials and external conditions, such as a laser field. This code allows one to lead with a set of anisotropic functions with different exponents for each space direction.

In the present work we shall use our code to study the energy spectrum of two electrons confined by a 3D anisotropic potential representing a 3D-DQD. We have adopted as confining potential a combination of the quartic potential $V(x,y)$ \cite{DiVincenzo-PRB59-2070-99, CRC-GJ-JAP94-2579-03}, for the $xy$ plane, with a parabolic potential on the $z-$direction \cite{Tarucha-Sci278-1788-97, Manninen-RMP74-1283-02}. We shall discuss the confinement of the electrons in the $xy$ plane as a function of the characteristic parameters of the system: the laser intensity,  the inter-dot distance, and the strengths of the potential along the $z-$ direction. 

Throughout the paper the computations were done in atomic units (au), more common in atomic-molecular calculations, whereas the results were expressed in meV and nm which are more tangible in nanoscale.

\section{The theoretical approach}
 We want to solve the time independent Schr\"odinger equation for a system of $N$ electrons submitted to an arbitrary potential $\hat{V}(x,y,z)$ whose Hamiltonian is written as:
\begin{eqnarray}
	\hat H =\sum_i^N\hat{O}_1(\vec r_i)+\sum_i^N\sum_{j<i}^N\hat{O}_2(\vec r_i,\vec r_j),
	\label{hamiltoniano}
\end{eqnarray}
where 
\begin{eqnarray}
\hat{O}_1(\vec r_i) = -\frac{1}{2m^*_c}\vec\nabla^2_i+ \hat{V}(x_i,y_i,z_i) ,
\end{eqnarray}
and
\begin{eqnarray}
	\hat{O}_2(\vec r_i,\vec r_j)=\frac{1}{\kappa\vert\vec r_i-\vec r_j\vert}.
\end{eqnarray}
The parameters  $\kappa$  and $m^*_c$ are respectively the static dielectric constant and the electron renormalized effective mass,  allowing us to considered general conditions  not necessarily in the vacuum.

\begin{figure}
\begin{center}
\includegraphics[scale = 0.35]{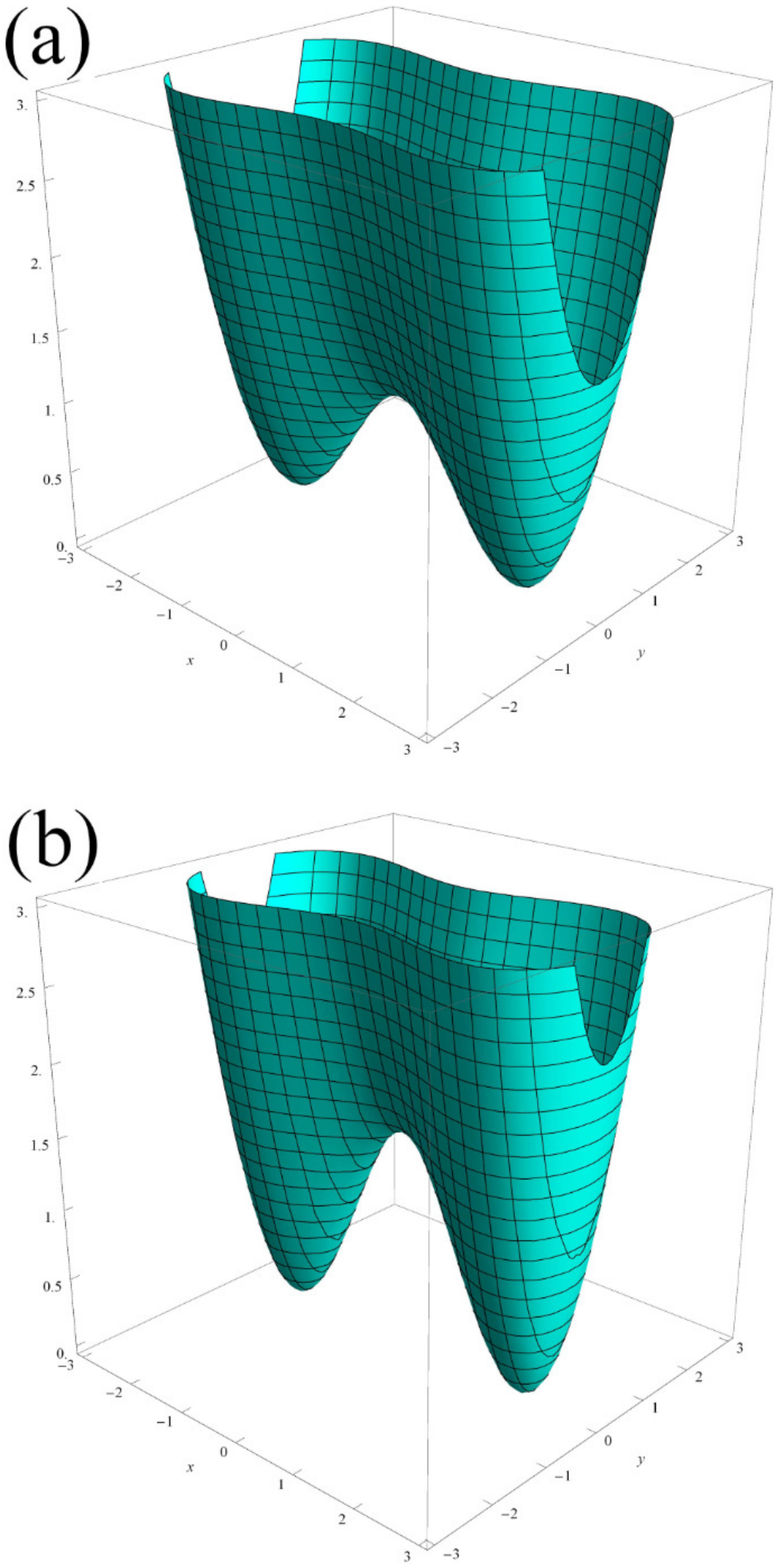}
\end{center}
\vspace{-1.8cm}\caption{3D visualization of the confining potential $\hat{V}(x,y,z)$ (Eq.~\ref{Vq}), in the $z=0$ plane. It is shown the effect of the effective mass for the case where $\omega_x = \omega_y$. For this pictorial visualization we have adopted a mass arbitrary unit such that in panel (a) $\hat{V}(x,y,0)$ corresponds to $m^*_c =1$, whereas in panel (b) corresponds to $m^*_c =1.5$ leading to deeper wells at $ x =\pm a$. }
\label{Quartic potential}
\end{figure}

In the present work we are interested in studying the electronic structure of a system composed of two electrons confined in a 3D CQD, whose potential is expressed as
\begin{eqnarray}
\hat{V}(x,y,z) = \frac{m^*_c}{2}\left[\frac{\omega_x^2}{4a^2}\left(x^2-a^2\right)^2+\omega^2_y y^2+\omega^2_z z^2\right].
\label{Vq}
\end{eqnarray}
The $xy$ dependence is modeled by a quartic potential $V(x,y)$\cite{DiVincenzo-PRB59-2070-99, CRC-GJ-JAP94-2579-03}. Observe that the advantage of using the quartic potential, in modeling the double quantum dot, consists in controling the size of the inter-dot barrier with the laser intensity without the necessity of changing any other parameter, see Fig.~\ref{Quartic potential}.  In the limit of inter-dot distance, $a\gg a^*_B$ where $a^*_B=\sqrt{1/(m^*_c\omega_x)}$, the potential splits into two harmonic wells of frequency $\omega_x$ and $\omega_y$ along $x$ and $y$, respectively. In the direction $z$ we assume an harmonic potential with frequency $\omega_z$, which can be chosen for instance to simulate a 2D double quantum dot by setting $\omega_z \gg \omega_x \mbox{ and } \omega_y$.

The electronic properties of free systems or confining potential in the study of quantum dots can be obtained due to the flexibility  of our program which can take into account the anisotropy of the potential on the basis employed \cite{olavo2016}. In addition, one can use a different  effective electronic mass $m^*_c$, once the laser is present through the electron renormalized effective mass, and/or  change the environment in which they evolve via the $\kappa$ parameter (see Ref.~\cite{CRC-GJ-JAP94-2579-03}).

The validity of the renormalized effective mass is discussed in detail in several works \cite{brandi1,brandi2,brandi3}. Briefly, the electronic band structure of the semiconductor is modeled by a two-parabolic, isotropic band in the $\vec{k}\cdot\vec{p}$ approximation \cite{callaway}. To incorporate the laser field into an effective mass formalism (renormalized effective mass approximation) the dressed atom \cite{Cohen} approach is extended to include a dispersion relation through the two-band model (dressed band approximation), the eigenvalue problem for the dressed bands is solved analytically and a $k$ expansion is performed.  According to this model the renormalized effective mass of the conduction band ($m^*_c$) is given by
\begin{equation}
\frac {1}{m^*_c}=\frac {1}{2M}\left[ 1+ \frac{M}{\mu} \frac{\left(
\frac{2\Lambda_0^2+\delta\Lambda_1}{\Lambda_1} \right) \left[
1-\frac{2\Lambda_0^2}{\Lambda_1^2} \left(1+\frac{2\Lambda_1}{E_g}
\right)\right] - \frac{4\Lambda _0^2}{E_g} }{\sqrt{4\Lambda_0^2+\left(
\frac{2\Lambda_0^2+\delta\Lambda_1}{\Lambda_1} \right)
^2}}\right] \label{massa}
\end{equation}
where $E_g$ is the energy gap, $ 1/M= 1/{m_c}+1/{m_v}$, $1/{\mu} = 1/{m_c}-1/{m_v}$,  and $m_c(m_v)$ is the undressed  effective mass associated to the conduction (valence) band: 
\begin{equation}
\frac{1}{m_c}=1 + \frac{2p^2}{Eg} \mbox{ and } \frac{1}{m_v}=1 - \frac{2p^2}{Eg},
\end{equation}
which leads to $m_c \approx 0.067$ and $m_v \approx -0,077$ for GaAs.
We have also defined the laser detuning parameter $\delta =E_g-\hbar\Omega$ and $\Lambda_1=2E_g-\delta$ and $\Lambda_0 = \left[ \left( 2I/I_c
\right) 7.02 \times E_g^2\right] ^{\frac 12}$. In the expression of $\Lambda_0$, $I_c$ is a critical intensity defined in Ref.~\cite{brandi1}, whose value for GaAs is $I_c\approx 5\times 10^{13}W/cm^2$. We have only considered the case of $\delta/E_g = 0.05$ (see Fig.1 of Ref.\cite{CRC-GJ-JAP94-2579-03}), and we have taken the range of intensity from $I/I_c = 0$ to  $10 \times 10^{-5}$. For this range the electron effective mass is displayed in the table~\ref{m_x_I}.

\begin{table}[h!]
\begin{center}
\caption{Electron effective mass as function of the laser field intensity. For details see the text.}
\label{m_x_I}
\begin{tabular}{|c|c|} 
\hline
 $(I/I_c)\times 10^{-5}$ & $m^*_c/m_c$ \\
\hline
0 & 1 \\ \hline
1. & 1.11025 \\ \hline
2. & 1.21163 \\ \hline
3. & 1.30639 \\ \hline
4. & 1.39599 \\ \hline
5. & 1.48149 \\ \hline
6. & 1.56364 \\ \hline
7. & 1.64304 \\ \hline
8. & 1.72013 \\ \hline
9. & 1.79527 \\ \hline
10. & 1.86877 \\ \hline
\end{tabular}
\end{center}
\end{table}

The solution of Eq.~(\ref{hamiltoniano}), $\Phi$, was obtained by a Full-CI method and is written as
\begin{equation}
\Phi=\sum_{i=1}^{N_\textrm{CSF}}C_i^{\textrm{CSF}}\Psi_i^{\textrm{CSF}}
\end{equation}
where $N_\textrm{CSF}$ is the number of configuration state functions (CSF) and $C_i^\textrm{CSF}$ represent the coefficient of a given CSF. On the other hand, a CSF is constitute of Slater determinants, {\em i.e.},
\begin{eqnarray}
\Psi_i^\textrm{CSF}=\sum_{i_1=1}^{\textrm{Ndet}_i}C_{i_1}^\textrm{det}\textrm{det}(i,i_1),
\end{eqnarray}
where $\textrm{det}(i,i_1)$ is the $i_1^{th}$ determinant of the $i^{th}$ CSF. As $[\hat{H},\hat{S}^2]=0$ and $[\hat{H},\hat{S}_z]=0$, $\Phi$ should be eigenfunction of $\hat{S}^2$ e $\hat{S}_z$.

\section{Setting the bases}

As mentioned in the introduction, we use a computational code to study the energy spectrum of two electrons confined in a 3D anisotropic potential. In order to employ it, we have to establish anisotropic orbitals as the atomic basis set. Similar to what was done in~\cite{olavo2016}, we have chosen a basis set composed of the Cartesian anisotropic Gaussian-type orbitals (c-aniGTO) centred in the position $\vec R=(X, Y, Z)$ which, apart a normalization constant, are  given by:
\begin{eqnarray}
&&g_\mu(\vec r-\vec R ,\zeta)=
(x-X )^{n_x}
(y-Y )^{n_y}
(z-Z )^{n_z}\times\nonumber\\
&&\exp\left[
-\zeta_x(x-X )^2
-\zeta_y(y-Y )^2
-\zeta_z(z-Z )^2
\right]
\label{gaussianxyz}
\end{eqnarray}
where one has the possibility of providing different exponents $\zeta_x$, $\zeta_y$ and $\zeta_z$ according to the problem analyzed and $\mu$ stands for $(n_x,n_y,n_z)$. In addition, in analogy to the standard convention for the atomic case, we shall  classify the orbitals as $s$-, $p$-, $d$-,... type according to $n=n_x + n_y + n_z = 0, 1, 2,...$, respectively.

Since the potential $V(x,y,z)$ along the $y$ and $z$ direction has the same form of the potential used in previous work~\cite{olavo2016}, the same two types of exponents have been considered in those directions: 
\begin{equation}
\zeta_i^{(1)}=\frac{m^*_c \omega_i}{2} \mbox{ and } \zeta_i^{(2)} = \frac{3}{2}\zeta_i^{(1)},
\end{equation}
where $i$ stands for $y$ and $z$. 

On the other hand, the first type of exponent in the $x$ direction, $\zeta_x^{(1)}$, has been obtained by a variational method minimizing the following functional,
\begin{eqnarray}
E(\zeta_x^{(1)})=\frac{\displaystyle\int_{-\infty}^{\infty}dx\phi_{\pm}^\ast(x,\zeta_x^{(1)})\hat O(x) \phi_{\pm}(x,\zeta_x^{(1)})}
{\displaystyle\int_{-\infty}^{\infty}dx\phi_{\pm}^\ast(x,\zeta_x^{(1)})\phi_{\pm}(x,\zeta_x^{(1)})},
\label{pvar_x}
\end{eqnarray}
where $\hat O(x) = \left[-\frac{1}{2m^*_c}\frac{d^2}{dx^2}+\frac{m^*_c\omega_x^2}{8a^2}(x^2-a^2)^2\right]$. 

The procedure to obtain this exponent was also employed in \cite{olavo2016} and is explained in its section 3. 
Once the potential displayed in the operator $\hat O(x)$ has minima in $x=\pm a$ and $y=z=0$, the functions $\phi_{\pm}(x,\zeta_x^{(1)})$ are taken as linear combination of the functions $g(\vec r-\vec R)$ centered at the same points.
This means that they correspond to $x$-direction molecular orbitals given by
\begin{eqnarray}
\phi_{\pm}(x,\zeta_x^{(1)})\hspace{-.1cm}&=& (x-a)^{n_x}e^{-\zeta_x^{(1)}\left(x-a\right)^2}  \nonumber \\
&& \pm (x+a)^{n_x}e^{-\zeta_x^{(1)}\left(x+a\right)^2} 
\label{sax}
\end{eqnarray}
However, we have observed that the function $\phi_{+}(x,\zeta_x^{(1)})$ provides lower values for the energy than the one obtained with $\phi_{-}(x,\zeta_x^{(1)})$.

Finally, the second type of the exponent was chosen similarly as the second type of the $y$ and $z$ exponents, namely $\zeta_x^{(2)} = 3\zeta_x^{(1)}/2$. Observe that due to the minimizing process the $\zeta_x^{(1,2)}$ exponents will depend on $n_x$.

Now, considering the excitations levels, given by $(n_x,n_y,n_z)$, as we are interested in confining only along the $z$ direction, we shall use larger values for the $\omega_z$. Consequently we expect few excitation in this direction, {\em i.e.}, we shall take only $n_z = 0, 1$, whereas in the plane $xy$ we will consider larger values: $n_x, n_y = 0, 1, 2,...$

The following results were obtained with a basis of 40 functions (2s2p2d) in each  center, with 820 (780) CSF's and 1600 (780) determinants for the singlet (triplet) states.

\section{Results and Discussion}

In the following we take $\omega_x=\omega_y=0.000111$ according to Ref.\cite{CRC-GJ-JAP94-2579-03} corresponding to a  confinement potential of 3~meV. As a typical value for the static  dielectric constant in GaAs, we consider $\kappa=13.6$. Besides, placing the coordinates origin in the middle of the dots, we consider two different values of the inter-dot distance $d=2a$: $a=270 a_0$ (14.3 nm) and 400$a_0$ (21.2 nm). We analyse three different confinement regimes, along the $z-$direction, whose strength is given by $\omega_z$.

\begin{figure}[h]
\begin{center}
\includegraphics[scale = 0.25]{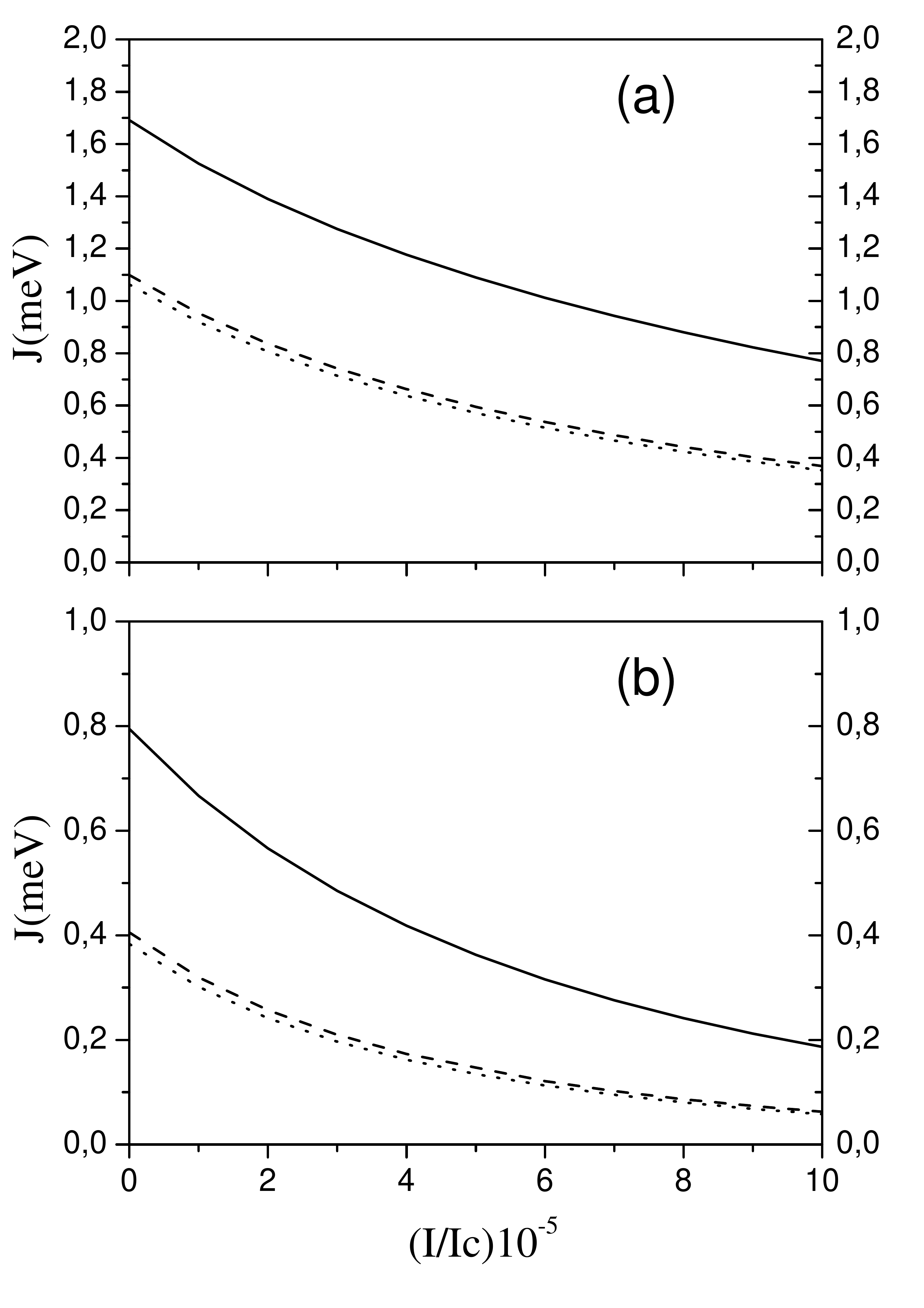}
\end{center}
\vspace{-.6cm}
\caption{Exchange coupling $J$ as function of the normalized laser intensity ($I/I_c$) for $\omega_z=0.000111$ (solid line), $\omega_z=0.0111$ (dashed line), and $\omega_z=0.111$ (dotted line). (a) Inter-dot distance $2a=540 a_0$ and (b) $2a=800$.}
\label{J=(T-S)}
\end{figure}

In Fig.~\ref{J=(T-S)} it is displayed the exchange coupling parameter ($J$) as function of the laser intensity. The parameter $J$ is defined as the energy difference between the first triplet and singlet states ($J=E_T - E_S$).

\begin{figure}[h]
\begin{center}
\includegraphics[scale = 0.25]{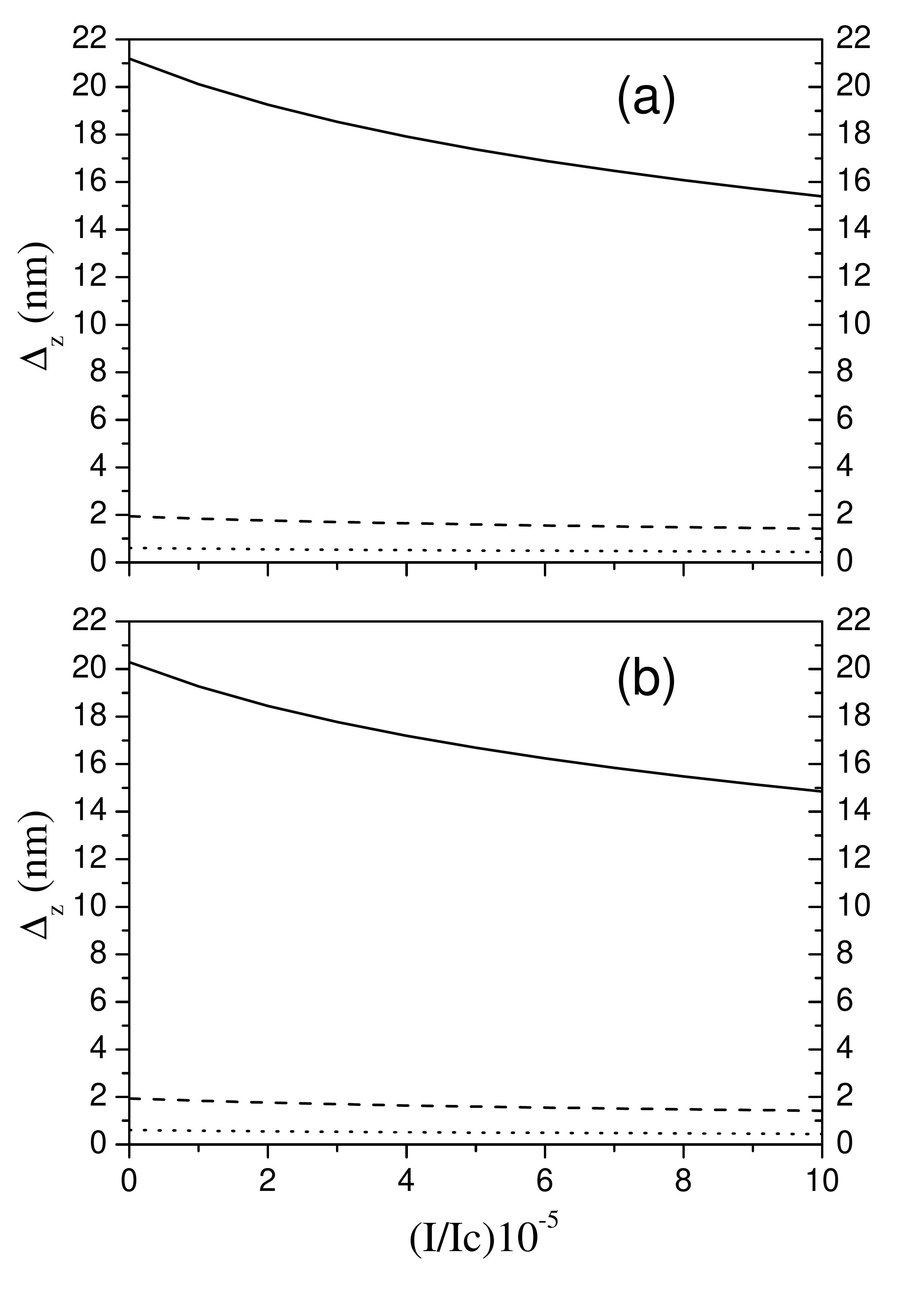}
\end{center}
\vspace{-.6cm}
\caption{Variance $\Delta _z$ as function of the normalized laser intensity ($I/I_c$) for $\omega_z=0.000111$ (solid line), $\omega_z=0.0111$ (dashed line), and $\omega_z=0.111$ (dotted line). (a) Inter-dot distance $2a=540 a_0$ and (b) $2a=800$.}
\label{Var_z}
\end{figure}

The confinement or compression in the $z-$direction is characterized when $\omega _z \gg \omega_x, \omega_y$ in Eq.~(\ref{Vq}). In Ref.~\cite{Diercksen10},  the value $\omega _z = 100\times\omega_x$ was sufficient to consider the electrons strongly compressed along the $z-$direction. In the present work, we have used as a confinement criterion in the $z-$direction the behavior of the root-mean-square of $z$ ($\Delta _z$) of the wave function defined as: 
\begin{equation}
\Delta _z=\sqrt{\langle z^2\rangle-\langle z\rangle^2}
\label{Dz}
\end{equation}
as a function of $\omega_z$. Indeed, Fig.~\ref{Var_z} confirm the confinement condition of Diercksen {\it et al}~\cite{Diercksen10} for the first singlet state. By observing the behavior of $J$ (Fig.~\ref{J=(T-S)}), one sees that there is a clear difference from $\omega_z=0.000111$ to $0.0111$, whereas from $\omega_z=0.0111$ to $0.111$ barely has any difference.

Next, we present the behavior of the electrons localization along the inter-dot direction ($x-$axis) by analysing the double-occupation probability in one of the dots. To do so we look at density function $\rho(x_1,x_2)$ defined as:
\begin{eqnarray}
\rho(x_1,x_2)=\int d\omega_1d\omega_2 dy_1dy_2 dz_1dz_2  \vert\Phi\vert^2, 
\label{1}
\end{eqnarray}
where $\Phi=\Phi(\vec r_1, \vec r_2,\omega_1,\omega_2)$, with  $\omega_1$ e $\omega_2$ representing the spin coordinates of the two electrons, and $\vec r_1=(x_1,y_1,z_1)$ and $\vec r_2=(x_2,y_2,z_2)$ their spatial coordinates.

\begin{figure}[h]
\begin{center}
\includegraphics[scale = 0.4]{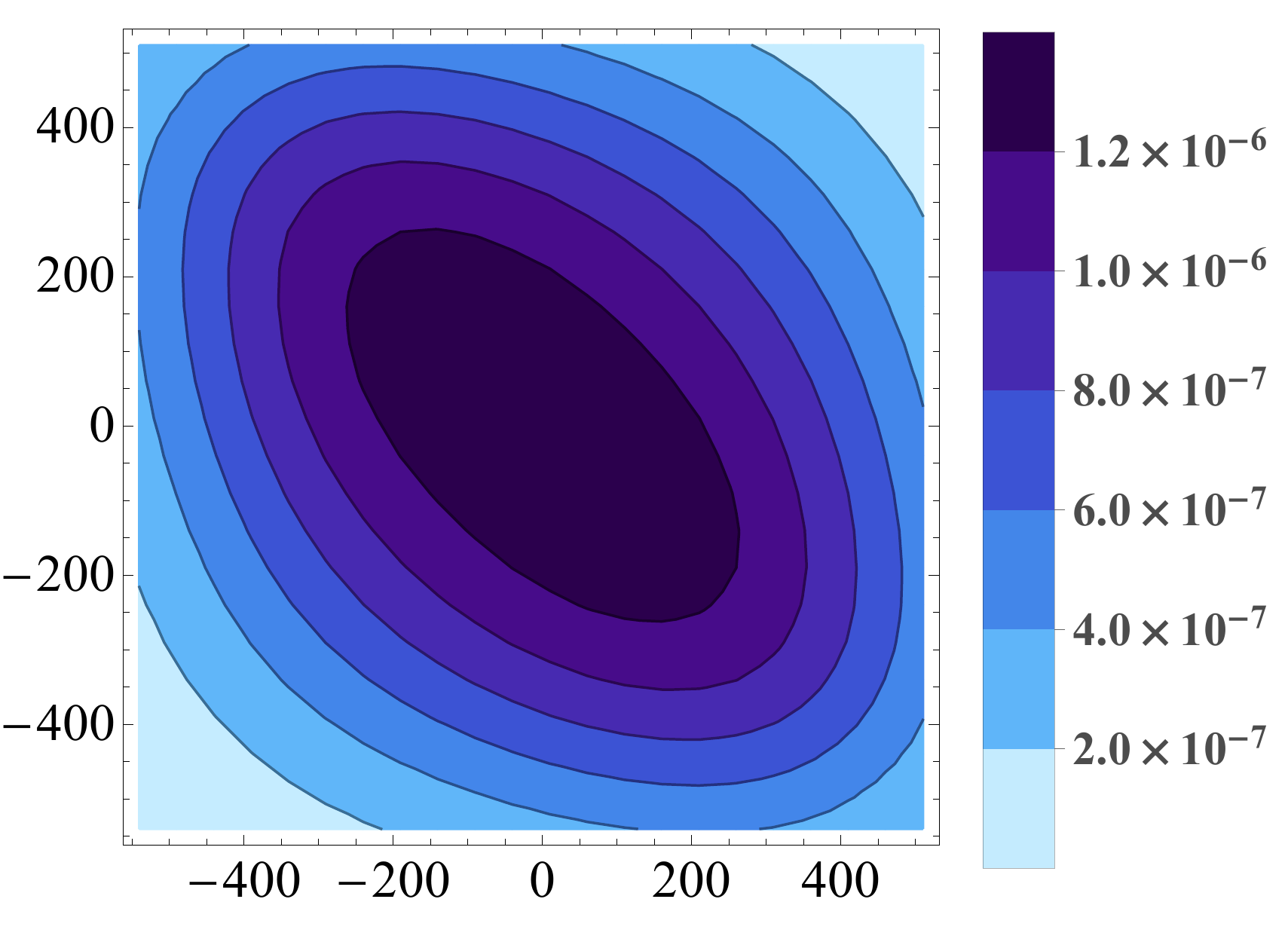}
\end{center}
\vspace{-.6cm}
\caption{Level curves of $\rho(x_1,x_2)$ for an inter-dot distance $2a=540$, $\omega_z=0.000111$ and $m^*_c/m_c=1.0$.}
\label{dens_mc-1_a-270_wz-000111}
\end{figure}

\begin{figure}[h]
\begin{center}
\includegraphics[scale = 0.4]{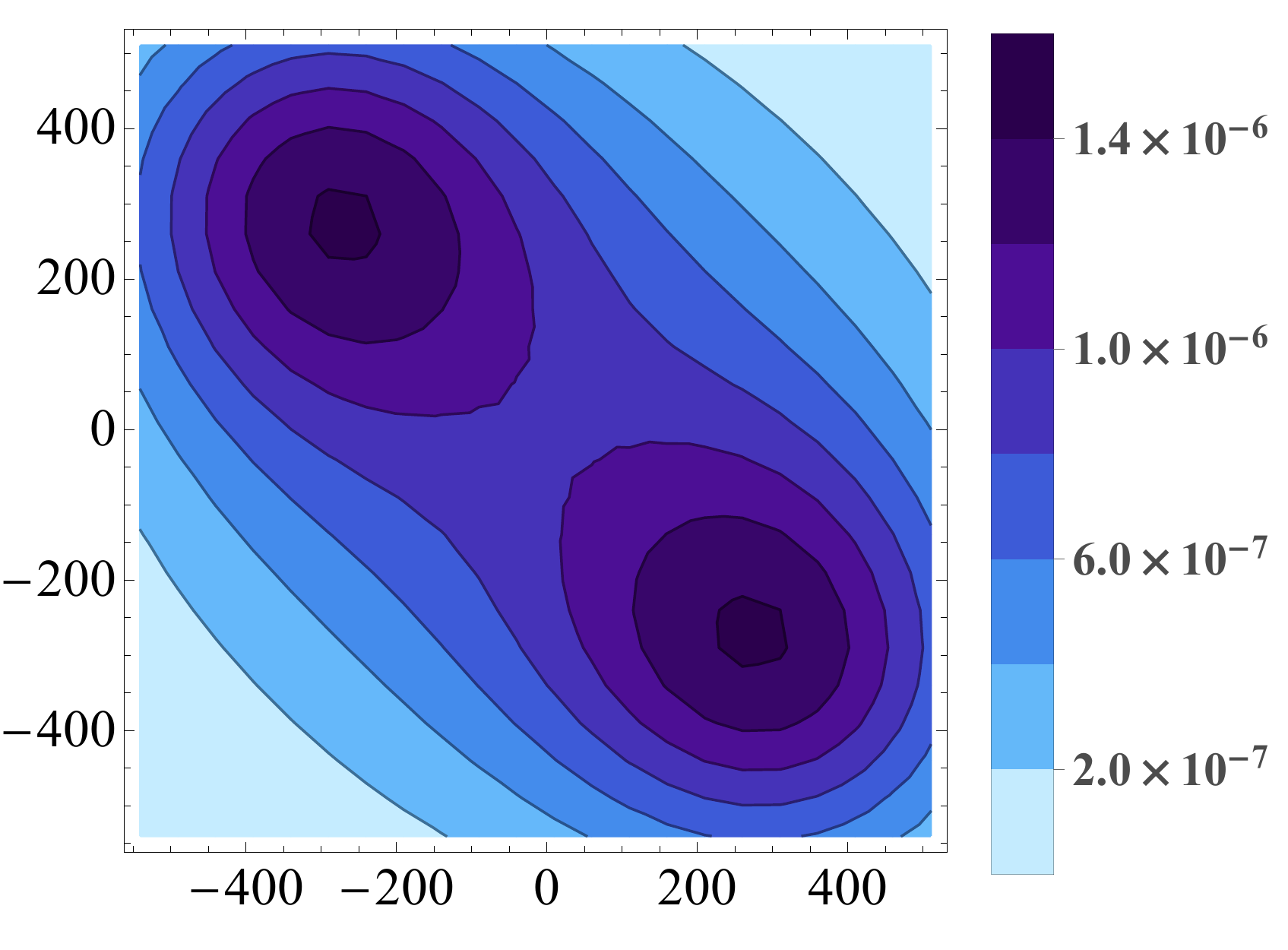}
\end{center}
\vspace{-.6cm}
\caption{Level curves of $\rho(x_1,x_2)$ for an inter-dot distance $2a=540$, $\omega_z=1.11$ and $m^*_c/m_c=1.0$.}
\label{dens_mc-1_a270_wz-111}
\end{figure}

\begin{figure}[h]
\begin{center}
\includegraphics[scale = 0.4]{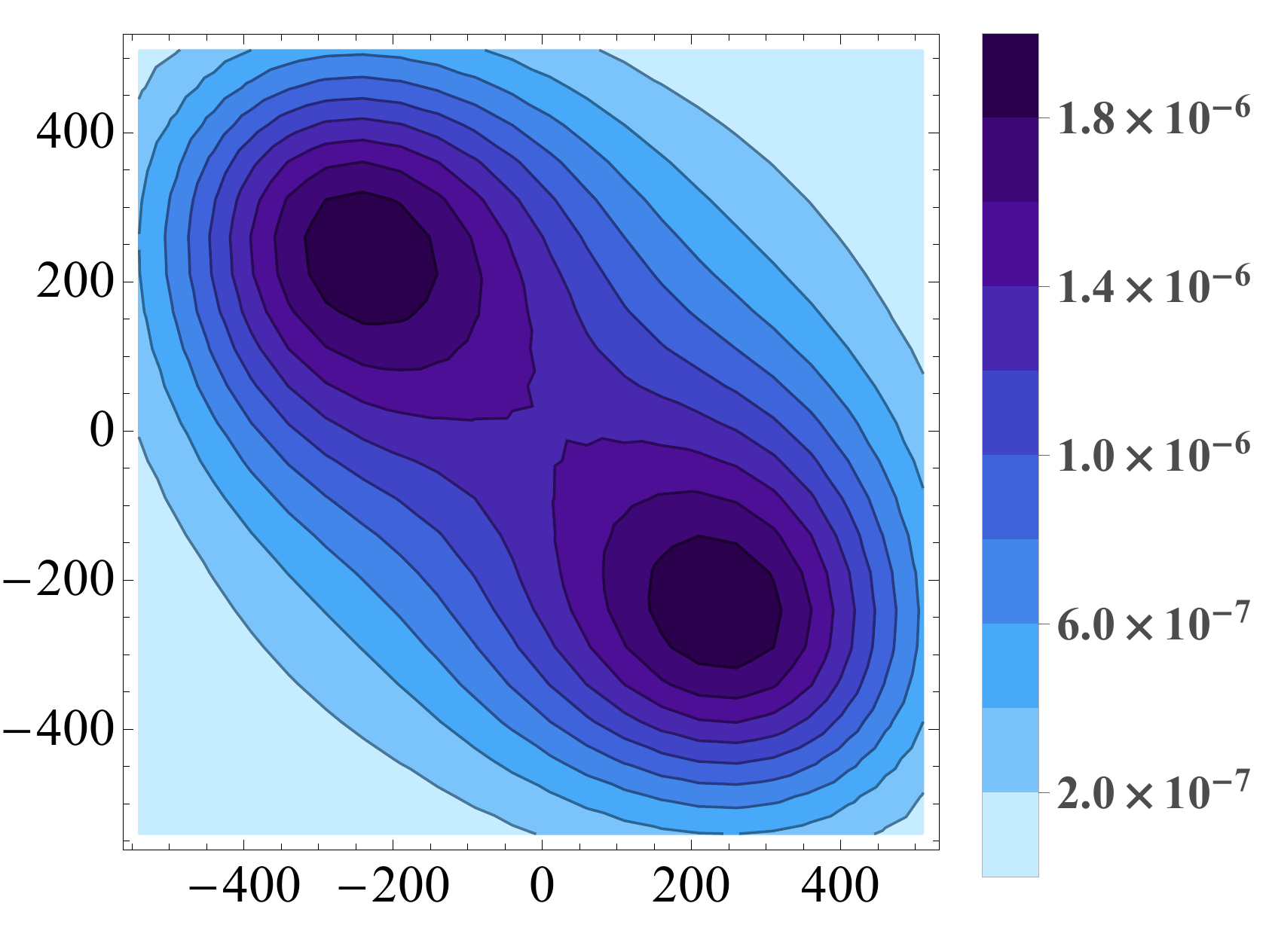}
\end{center}
\vspace{-.6cm}
\caption{Level curves of $\rho(x_1,x_2)$ for an inter-dot distance $2a=540$, $\omega_z=0.000111$ and $m^*_c/m_c=1.86877$.}
\label{dens_mc-18_a270_wz-000111}
\end{figure}

\begin{figure}[]
\begin{center}
\includegraphics[scale = 0.4]{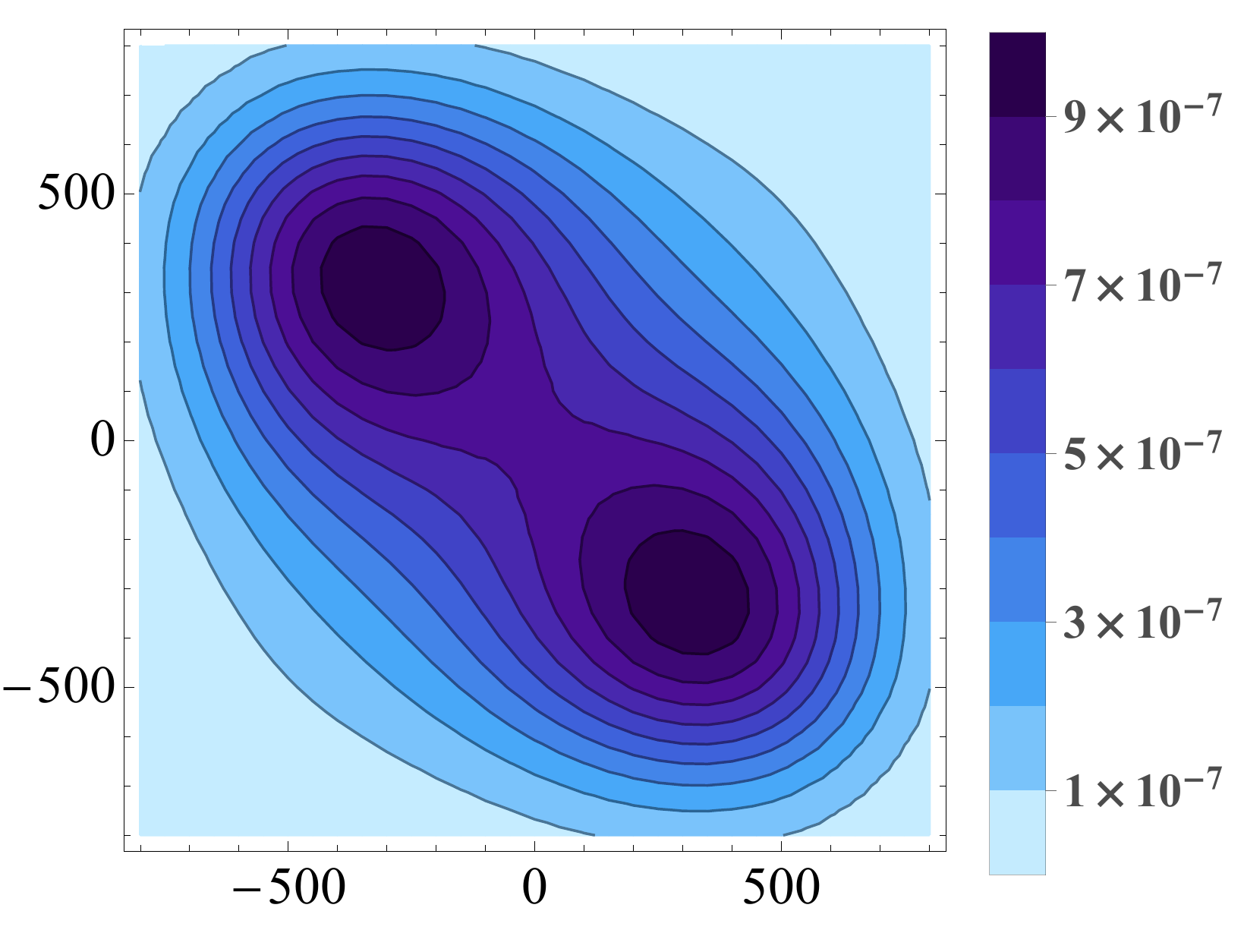}
\end{center}
\vspace{-.6cm}
\caption{Level curves of $\rho(x_1,x_2)$ for an inter-dot distance $2a=800$, $\omega_z=0.000111$ and $m^*_c/m_c=1.0$.}
\label{dens_mc-1_a-400_wz-000111}
\end{figure}

\begin{figure}[h]
\begin{center}
\includegraphics[scale = 0.4]{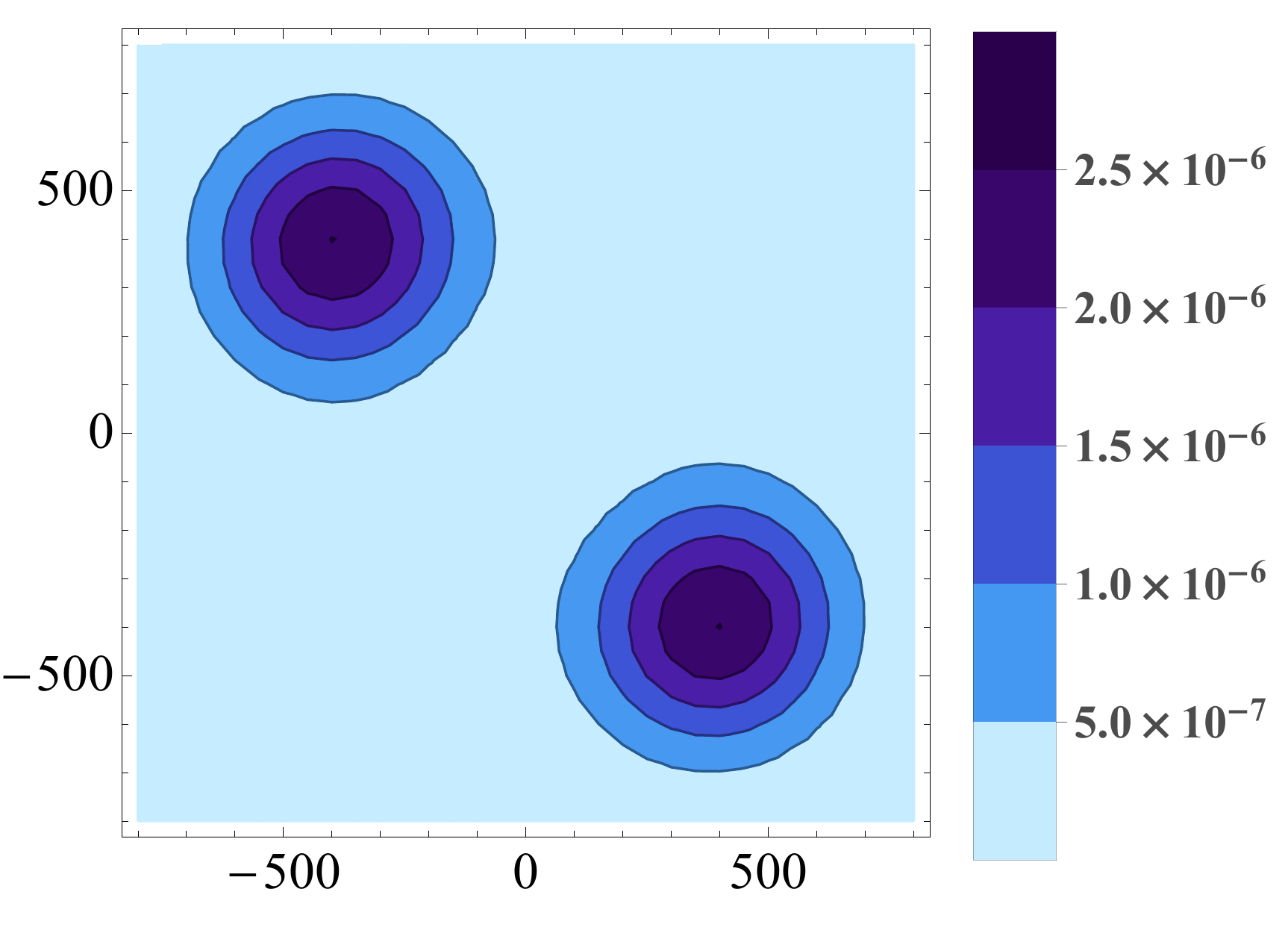}
\end{center}
\vspace{-.6cm}
\caption{Level curves of $\rho(x_1,x_2)$ for an inter-dot distance $2a=800$, $\omega_z=1.11$ and $m^*_c/m_c=1.86877$.}
\label{dens_mc-18_a-400_wz-111}
\end{figure}

In Figs.~\ref{dens_mc-1_a-270_wz-000111}~--~\ref{dens_mc-18_a-400_wz-111} are displayed the contour plots of $\rho(x_1,x_2)$ for different conditions. We have only considered the system in its fundamental state, the singlet, to analyse the electrons spatial positioning along the $x-$axis.  This choice is based in what is observed in Fig.~\ref{J=(T-S)}, where $J=E_T-E_S$ is always positive. The graphic horizontal and vertical axes, $x_1$ and $x_2$, respectively, correspond to the position of electron 1 and 2 along the $x-$axis; once the electrons are undistinguishable, we expect to have reflection symmetry in respect to the diagonal line $x_1=x_2$.

We analyze the double-occupation  as a function of the laser field intensity, via the effective mass $m^*_c/m_c$, the distance $a$ and the $z-$axis confinement parameter $\omega_z$. 

Thus, in Fig.~\ref{dens_mc-1_a-270_wz-000111} it is shown $\rho(x_1,x_2)$ for  $m^*_c/m_c=1$, $a=270 a_0$ and $\omega_z=0.000111$. One observes that the probability of finding both electrons in the middle of the two dots is maximum; for small values of $x_1$ and $x_2$ simultaneously, one obtains the largest values of  $\rho(x_1,x_2)$. At the same time, there is a considerable chance of finding them in the same dot $\rho(270,270)=\rho(-270,-270) \approx \frac{1}{2}\rho(\sim 0, \sim0)$.

Fig.~\ref{dens_mc-1_a270_wz-111} shows the behaviour of $\rho(x_1,x_2)$ for the same parameters $m^*_c/m_c$ and $a$, but under an extreme large confinement in the $z-$direction ($\omega_z=1.11$). Now, one observes that the maximum probability occurs at $\sim (270, -270)$, and at the corresponding symmetrical place $\sim (-270, 270)$. This means that under strong $z-$confinement the electrons drain from the inter-dots region to the dots; consequently the probability of finding both electrons at the same dot becomes very low.

Although we have analysed the effect of confinement up to a strength $\omega_z=0.111$ in Figs.~\ref{J=(T-S)}, here we have chosen an extreme confinement condition, corresponding to $\omega_z=1.11$, in order to compare with the regime of intense laser field, which is displayed in Fig.~\ref{dens_mc-18_a270_wz-000111}, where one can see the confinement property of the laser field. Now one can observe that the behaviour of $\rho(x_1,x_2)$, for the same  parameters  $a$ and $\omega_z$ as in Fig~\ref{dens_mc-1_a-270_wz-000111} but with a higher mass $m^*_c/m_c=1.86877$, is similar to the one of  Fig.~\ref{dens_mc-1_a270_wz-111}.

Now, let us  look at the confinement property of the distance as in Fig.~\ref{dens_mc-1_a-400_wz-000111}. We observe that the behaviour of $\rho(x_1,x_2)$ for    $a=400 a_0$ and $\omega_z=0.000111$ is similar to the one observe in    Fig.~\ref{dens_mc-18_a270_wz-000111}. 

Finally, the  confinement properties  of the distance and laser field intensity are observed in  Fig.~\ref{dens_mc-18_a-400_wz-111}. We observe that the behaviour of $\rho(x_1,x_2)$ for    $a=400 a_0$ and $\omega_z=1.11$ and $m^*_c/m_c=1.86877$ characterizes a situation where the electrons are localized in the opposite dots. 

It is worth mentioning that the behaviour of the density for the first triplet state was also calculated. It indicates that the electrons tend to stay away from each other, each in a dot, for all considered conditions, as it was expected. Thus they are not presented.

\section{CONCLUSIONS}
In this work we have analysed the confinement of the electrons in a coupled quantum dot. We have confirmed a criterion established in the literature concerning the confinement in the $z-$direction, analysing the exchange coupling $J$ and the dispersion of the electrons along the $z$ axis through the electrons position variance $\Delta_z$. In addition, we have presented another way of confining the electrons by applying a laser field. The advantage of using laser field is that one can vary the confinement in a simple manner, in contrast to others manners which involve the parameters $a$ (the inter-dot distance) and $\omega_z$ (connected to the potential profile along the $z-$direction) both constant or, at least, difficult to manage or vary. In order to establish that, we have performed calculations using a Full-CI wave functions to obtain information about the double-occupation of the electrons.

\section*{Acknowledgments}
This work was partially supported by the Brazilian agencies CNPq,  CAPES, FAPESB and FAPERJ.
\\

\end{document}